%% file: main.tex
\documentclass[%
 superscriptaddress,
 reprint,
 showpacs,preprintnumbers,
 nofootinbib,
 amsmath,amssymb,
 aps,
 prd,
 floatfix,
]{revtex4-1}

\usepackage{amsmath,amssymb,amsfonts}
\usepackage{graphicx}
\usepackage{hhline}
\usepackage{bm}
\usepackage{amsmath}
\usepackage{epsfig}
\usepackage{graphicx}
\usepackage{xcolor}
\usepackage{siunitx}
\usepackage{comment}
\usepackage{tabularx}
\usepackage{rotating}
\usepackage{pstricks}
\usepackage{subcaption}
\usepackage{footnote}
\usepackage{hyperref}
\usepackage{ulem}

\DeclareSIUnit{\clight}{\mathit{c}}

\begin{document}
\title{Implementation of the Martini-Ericson-Chanfray-Marteau RPA-based neutrino and antineutrino cross-section model in the GENIE neutrino event generator}

\author{L. Russo}
\email{lrusso@lpnhe.in2p3.fr}
\affiliation{Sorbonne Universit\'e, CNRS/IN2P3, Laboratoire
de Physique Nucl\'eaire et de Hautes Energies (LPNHE), Paris, France}
\author{M. Martini}
\email{mmartini@lpnhe.in2p3.fr}
\affiliation{IPSA-DRII,  63 boulevard de Brandebourg, 94200 Ivry-sur-Seine, France}
\affiliation{Sorbonne Universit\'e, CNRS/IN2P3, Laboratoire
de Physique Nucl\'eaire et de Hautes Energies (LPNHE), Paris, France}
\author{S. Dolan}
\affiliation{CERN, European Organization for Nuclear Research, Geneva, Switzerland}
\author{L. Munteanu}
\affiliation{CERN, European Organization for Nuclear Research, Geneva, Switzerland}
\author{B. Popov}
\affiliation{Sorbonne Universit\'e, CNRS/IN2P3, Laboratoire
de Physique Nucl\'eaire et de Hautes Energies (LPNHE), Paris, France}
\author{C. Giganti}
\affiliation{Sorbonne Universit\'e, CNRS/IN2P3, Laboratoire
de Physique Nucl\'eaire et de Hautes Energies (LPNHE), Paris, France}

\begin{abstract}
\input{0_abstract}
\end{abstract}

\maketitle
\section{Introduction}
\label{sec:intro}
\input{1_intro}

\section{The theoretical model}
\label{sec:theo}
\input{2_theomodel}

\section{Implementation in GENIE and validation}
\label{sec:genie}
\input{3a_implementation}
\input{3b_limitations}

\section{Comparison with measurements}
\label{sec:comp_exp}

\input{4_datacomparison}

\section{Summary and Conclusion}
\label{sec:conclusions}
\input{5_conclusions}

\section*{Acknowledgments}
\input{6_acknoledgements}

\bibliography{main.bbl}

\appendix
\input{7_appendix}

\end{document}

%% file: 0_abstract.tex
We discuss the first implementation of the Martini-Ericson-Chanfray-Marteau random phase approximation-based (anti)neutrino cross-section model
for quasielastic (1p1h) and multinucleon (2p2h and 3p3h) excitations in the widely used GENIE neutrino event generator.
Validation steps are presented, in particular, through direct comparisons of GENIE cross-section output with original calculations performed by the authors of the model. Predictions for $^{12}$C, $^{16}$O, and $^{40}$Ar
are compared with some available T2K and MicroBooNE experimental measurements showing a reasonable agreement.

%% file: 1_intro.tex
The success of the next generation long-baseline accelerator-based neutrino oscillation experiments Hyper-Kamiokande~\cite{Hyper-Kamiokande:2018ofw} and DUNE~\cite{DUNE:2015lol} will depend on the reduction to the percent level of the systematic errors. Presently, the most important source of uncertainties is represented by the neutrino-nucleus cross sections ~\cite{Katori:2016yel,NuSTEC:2017hzk}, due to the difficulty in modeling nuclear dynamics, the nucleus being a quantum many-body system whose constituents interact through the strong force. To improve the knowledge of neutrino cross sections, a common and joint effort of experimental measurements, theoretical calculations and model implementations in Monte Carlo event generators is needed. These three aspects cannot be separated from each other and are entangled as three Borromean rings. Think, for example, of the neutrino energy reconstruction problem, the background subtraction for a peculiar experimental signal topology, or the modeling of the detector response.

In this paper, we present the implementation of the theoretical random phase approximation (RPA) based model of the Lyon group, described in detail in the article of Martini \textit{et al.}~\cite{Martini:2009uj} into the GENIE event generator~\cite{Andreopoulos:2009rq}. 
We then compare its predictions with different experimental measurements. This model 
allows a unified description of several scattering channels: 
quasielastic (QE), multinucleon emission and the coherent
and incoherent single pion production. 
Within this model, the debated MiniBooNE quasielasticlike cross section on carbon~\cite{AguilarArevalo:2010zc} has been explained for the first time~\cite{Martini:2009uj,Martini:2010ex} by stressing the crucial role of the multinucleon (also called 2p2h or npnh) excitation contributions which sizably increase the cross section. After this explanation, 2p2h excitations started to attract a lot of attention (for a review see, for example, Refs.~\cite{Katori:2016yel,NuSTEC:2017hzk}), first in the neutrino scattering community, then in the neutrino oscillation one. This growing interest is due to the central role of npnh, together with quasielastic processes, in the accurate reconstruction of the neutrino energy using the kinematic method employed by the T2K~\cite{Abe:2011ks} and Hyper-K experiments~\cite{Martini:2012fa,Martini:2012uc,Nieves:2012yz,Lalakulich:2012hs,Ankowski:2015jya} and, consequently, in the precise determination of the neutrino oscillation parameters. All widely-used event generators now include the multinucleon excitation channel, previously ignored before the MiniBooNE quasielasticlike results, into their packages. For a relatively recent state-of-the-art implementation in the different generators see, for example, Refs.~\cite{GENIE:2021npt,Hayato:2021heg,Golan:2012rfa,Buss:2011mx}, as well as the dedicated articles~\cite{Schwehr:2016pvn,Dolan:2019bxf,Bourguille:2020bvw,Prasad:2024gnv,Mosel:2023zek,Ankowski:2025umq}. This implementation effort is still ongoing, essentially for two reasons. First, various theoretical models present some differences in their predictions for this channel, whose complexity--stemming from the intrinsic difficulty of nuclear physics--makes it hard to treat without approximations. Second, the experimental neutrino community is moving toward more exclusive measurements based on detected hadronic variables, whereas most of the existing models were originally developed for inclusive cross sections in terms of leptonic observables.
This interest in the hadronic variables is driven by the improved sensitivity to outgoing low-energy protons and neutrons of the upgraded T2K near detector ND280~\cite{T2K:2019bbb} as well as by the present Short-Baseline Neutrino Program at Fermilab~\cite{MicroBooNE:2015bmn} and the future DUNE experiment~\cite{DUNE:2015lol}, both of which use liquid argon detectors with especially good hadronic sensitivity. The calorimetric energy estimator used there can also be problematically sensitive to the npnh effects (e.g. missing energy due to emitted neutrons or subthreshold protons).

Up until now, the model of Ref.~\cite{Martini:2009uj} has not been included in any generators by the implementation of the corresponding theoretical hadron tensor, but only employing tables of shared cross sections at different kinematics via a reweighting procedure~\cite{Nguyen:2022nmy}. The hadron tensor strategy is employed here for the first time for this model. For a detailed explanation of this strategy, see~Ref.~\cite{Schwehr:2016pvn} where it was developed as well as Ref.~\cite{Dolan:2019bxf}, where it is discussed in the context of the SuSAv2 model. Even though, in principle, all the channels described by this model can be implemented in the generators, in the present work we confine ourselves to considering only the quasielastic (also called 1p1h) and the multinucleon excitation channels (2p2h and 3p3h, collectively referred to as npnh), which have been the most investigated and employed in the neutrino analyses of accelerator-based experiments over the last fifteen years. For this purpose we follow the same protocol as the one adopted in Refs.~\cite{Dolan:2019bxf,Dolan:2021rdd} for the SuSAv2+ meson-exchange current (MEC)~\cite{Megias:2016fjk} (quasielastic + 2p2h) and Ghent continuum random-phase approximation (CRPA)~\cite{Pandey:2014tza} (giant resonances + quasielastic) models, respectively. 

The interest in implementing the Martini-Ericson-Chanfray-and Marteau model (hereafter referred to as the Martini \textit{et al.} model) relies on several aspects. First, as already mentioned, differences appear among the various models, and none is currently able to simultaneously describe all the available experimental measurements across the full neutrino energy range--from low (e.g., T2K) to high (e.g., MINERvA) energies--and across nuclear targets (most notably, $^{12}$C, $^{16}$O, and $^{40}$Ar). In the absence of such a solution, the span of the predictions from plausible models already implemented inside generators provides some guidance about, for example, the size of systematic errors associated with different processes (e.g., see the heavy use of this approach in Ref.~\cite{Wilkinson_2022}).
Second, in the Martini \textit{et al.} model, the different 2p2h contributions--namely the ones related to (i) nucleon-nucleon correlations, (ii) meson exchange currents, and (iii) their interference--as well as the 3p3h excitations, are separately calculated and available as distinct components of the hadron tensor. This allows for easier comparison between 2p2h models and, overall, could help avoid the risk of double counting (e.g., nucleon-nucleon correlations contributions) when combining different channels taken from different models (e.g., using a spectral function model, like that in Ref.~\cite{Benhar:2006nr}, for the quasielastic channel and Fermi gas-based models for 2p2h interactions). Finally, calculations for 
$^{12}$C, $^{16}$O and $^{40}$Ca 
are available, enabling the modeling of cross sections for carbon, water, and argon targets--the main targets of present and future accelerator-based neutrino experiments. 

Even if the Martini \textit{et al.} model allows also the treatment of neutral-current (NC) neutrino cross sections, and the nuclear responses are available also for the electron scattering, in the present work we focus exclusively on the GENIE implementation of the neutrino and antineutrino (1p1h and npnh) charged-current (CC) cross sections. 

The article is organized as follows. In Sec.~\ref{sec:theo}, we briefly recall the main aspects and achievements of the model presented in Ref.~\cite{Martini:2009uj}. In Sec.~\ref{sec:genie}, we describe the strategy for implementing this model in GENIE and its validation. In Sec.~\ref{sec:comp_exp}, we compare the model predictions with several sets of measurements, with the aim of illustrating the model's potential for describing different reaction channels, various final-state topologies, and different targets. Finally, in Sec.~\ref{sec:conclusions}, we draw our conclusions.

%% file: 2_theomodel.tex
The theoretical model is described in Ref.~\cite{Martini:2009uj}, to which we refer for all the details. It is based on nuclear response functions treated in the RPA on top of a local Fermi gas (LFG) calculation. 
This formalism belongs to a long tradition of investigations: first on electron scattering and pionic modes in nuclei (see, for example, Refs.~\cite{Delorme:1980dle,Alberico:1980vb,Alberico:1981sz,Alberico:1983zg,Stroth:1985ybp,Alberico:1985zz,Alberico:1986kn,PhysRevA.38.4832,Delorme:1989nh,Laktineh:1992kp,Chanfray:1993um,Alberico:1997jg}) and, later on, neutrino scattering~\cite{Delorme:1985ps,Marteau:1999kt, Marteau:1999jp}. The model of Ref.~\cite{Martini:2009uj} is the continuation and extension of these works. One of its strengths is that it allows a unified description of several scattering channels: quasielastic, multinucleon emission and the coherent and incoherent single pion production. As already mentioned, within this model, the debated MiniBooNE quasielasticlike cross section on carbon~\cite{AguilarArevalo:2010zc} was explained for the first time~\cite{Martini:2009uj,Martini:2010ex} by advocating the crucial role of the multinucleon excitations. 
In Ref.~\cite{Martini:2011wp}, an improvement of the original description of the quasielastic channel was made through the introduction of relativistic corrections. This relativistic description of the quasielastic channel is implemented in the present work. The Martini \textit{et al.} model has successfully reproduced many of the features of several muon-neutrino flux-integrated double differential cross-section measurements in the past: 
the MiniBooNE charged-current quasielasticlike neutrino~\cite{Martini:2011wp} and antineutrino~\cite{Martini:2013sha} measurements, as well as their combination~\cite{Ericson:2015cva}; the MiniBooNE CC1$\pi^+$ and the T2K CC inclusive measurements~\cite{Martini:2014dqa,Martini:2022ebk}; the T2K CC0$\pi$ for neutrino~\cite{Abe:2016tmq} and antineutrino measurements, as well as their combination~\cite{Abe:2020jbf}.
Single differential electron-neutrino T2K CC inclusive cross sections have been successfully reproduced as well~\cite{Martini:2016eec}. More recently, the model has been employed to investigate the MicroBooNE CC inclusive~\cite{Martini:2022ebk} and NC1$\pi^0$~\cite{Martini:2024mtx} cross sections on argon. The counterintuitive dominance of the $\nu_\mu$ differential cross sections over the $\nu_e$ ones in specific kinematical conditions--not only for quasielastic, but also for multinucleon excitations and single pion production--has been explained with simple arguments in Ref.~\cite{Martini:2023kem} using this model. 

Even though it has not yet been included in any Monte Carlo event generator via the implementation of the corresponding hadron tensor, the model has already been considered in many experimental neutrino cross-section measurement comparisons~\cite{Abe:2016tmq,Abe:2020jbf,NOvA:2020rbg,T2K:2023qjb} and oscillation analyses~\cite{T2K:2017rgv,T2K:2018rhz,T2K:2019bcf,T2K:2023smv,T2K:2024wfn}. In these experimental papers, the model is often referred to as  ``Martini \textit{et al.}'' or ``Lyon.'' The main reason for considering it is its larger cross-section predictions in the 2p2h sector (especially for neutrinos and, to a lesser extent, for antineutrinos)  compared to those of Ref.~\cite{Nieves:2011pp}, a similar theoretical approach, often referred to as ``Nieves \textit{et al.}'' or ``Valencia,'' which is already implemented in many generators. A very recent reexamination~\cite{Sobczyk:2024ecl} of the Nieves \textit{et al.} approach now leads to new results that are closer to that of Martini \textit{et al.}

Before concluding this section, for clarity and to make the work as self-contained as possible (without entering in too much detail already published elsewhere) we remind the reader of some general aspects of the neutrino-nucleus cross sections and of the formalism employed by Martini \textit{et al.} in their calculations. 

The double differential cross section for the charged-current reaction
\mbox{$ \nu_l \, (\bar{\nu}_l) + A \longrightarrow l^- \, (l^+) + X $}
is given by
\begin{equation}
\label{m_eq_1}
\frac{d^2 \sigma}{d\Omega_{k'} d \omega} =  
\frac{G_F^2 \cos^2\theta_C}{32\pi^2}\frac{|{\bf{k}}'|}{|{\bf{k}}|}L_{\mu\nu}W^{\mu\nu}({\bf{q}},\omega). 
\end{equation}
Here $d\Omega_{k'}$ is the differential solid angle in the direction specified
by the charged lepton momentum ${\bf{k}}'$ in the laboratory frame,
$\omega=E_\nu-E_l'$ is the energy transferred to the nucleus,
the zero component of the four-momentum transfer $q=k-k^\prime\equiv(\omega,{\bf{q}})$,
with $ k \equiv (E_\nu,{\bf{k}})$ and $ k^\prime \equiv (E_l',{\bf{k}}')$, being the initial and final lepton four momenta.
In Eq.~(\ref{m_eq_1}), $ G_F $ is the weak coupling constant,
$ \theta_C$ is the Cabibbo angle. 

$L_{\mu\nu}$ is the leptonic tensor
\begin{equation}
L_{\mu\nu} = 8 (k_\mu {k'}_\nu + k_\nu {k'}_\mu - g_{\mu\nu} k \cdot k' \mp i \varepsilon_{\mu\nu\alpha\beta} k^\alpha {k'}^\beta),
\end{equation}
and $W^{\mu\nu}({\bf{q}},\omega)$ is the hadronic tensor, a combination of nucleon form factors (that contain the information about the nucleon properties),
and nuclear response functions (that contain the information about the nuclear dynamics). The different components of the hadronic tensor can be combined allowing a reformulation of Eq.~(\ref{m_eq_1})
in terms of projections with respect to the momentum transfer direction.
The charged-current cross section is a linear combination of five contributions
\begin{eqnarray}
\label{m_eq_general}
\frac{d^2 \sigma}{d\Omega_{k'} d \omega} &=&  
\sigma_0\left[L_{00}W^{00} \, + \, L_{33}W^{33} \, + \, (L_{03}+L_{30})W^{03} \,\right.\nonumber \\ 
&+& \left. (L_{11}+L_{22})W^{11} \, \pm \, (L_{12}-L_{21})W^{12} \right],
\end{eqnarray}
where the kinematical factors come from the contraction with the leptonic tensor and
the plus (minus) sign applies to neutrinos (antineutrinos). We do not provide here the explicit expressions for the kinematical factors and for the nucleon form factors and nuclear responses entering in the hadronic tensors, as they can be found in Appendixes~A and B of Ref.~\cite{Martini:2009uj}.\footnote{There is a typo in the expression of $L_{12}-L_{21}$ in Eq.~(A4) of Ref.~\cite{Martini:2009uj}: the factor 8 should be replaced by the factor 16.} 

The hadronic tensor can be decomposed in terms of the following scattering channels:
\begin{equation}
\label{eq-dec_W_ph}
 W^{\mu\nu} = W^{\mu\nu}_{1p1h} + W^{\mu\nu}_{2p2h}  + W^{\mu\nu}_{3p3h} +W^{\mu\nu}_{1p1h1\pi} +W^{\mu\nu}_{1\pi\textrm{coh.}}
\end{equation}
that are available as subtables of tensor elements.
It can also be decomposed in terms of the different channels of particle-hole (nucleon $N$ or Delta $\Delta$) excitations
\begin{equation}
\label{eq-dec_W_ND}
 W^{\mu\nu} = W^{\mu\nu}_{NN} + W^{\mu\nu}_{N\Delta} + W^{\mu\nu}_{\Delta\Delta}, 
\end{equation}
which are particularly useful to classify the type of 2p2h contributions. 
It is important to remember that Eqs.~(\ref{m_eq_1}) and (\ref{m_eq_general})  
are totally general and, according to the decompositions of Eq.~(\ref{eq-dec_W_ph}),  apply to different excitations channels: 1p1h quasielastic, 2p2h, 3p3h (the three channels considered in the present work), and 1p1h 1$\pi$ production,
coherent $\pi$ production (the two other channels of the Martini \textit{et al.} model, whose Monte Carlo implementation is left for future works).

Since the present work focuses mainly on multinucleon excitations, we recall that in the model of Ref.~\cite{Martini:2009uj} the following contributions are included: 
the nucleon-nucleon correlations (denoted NN in Ref.~\cite{Martini:2009uj}); the 
$\Delta$-MEC contribution (denoted $\Delta\Delta$ in Ref~\cite{Martini:2009uj}); and the NN correlations--$\Delta$-MEC interference (denoted N$\Delta$ in Ref.~\cite{Martini:2009uj}). These contributions are separately available as subtable of the hadron tensor  
$W^{\mu\nu}_{2p2h}$, 
according to the decompositions of Eq.~(\ref{eq-dec_W_ND}) 
, which holds for the different scattering channels. 
We also recall that, as in all the works of the Lyon group starting from the end of Ref.~\cite{Martini:2009uj}, only the npnh evaluation called ``new'' in Ref.~\cite{Martini:2009uj} is considered; the ``old'' version corresponds to the one employed in Refs.~\cite{Marteau:1999kt, Marteau:1999jp}. This so-called new evaluation is based, for the NN and N$\Delta$ parts, on the microscopic calculations of Ref.~\cite{Alberico:1983zg}, aimed at determining the 2p2h contribution
to the isospin spin-transverse response. For the $\Delta\Delta$ part, the contributions related to the nonpionic decay of the $\Delta$ inside the nuclear medium--leading to 2p2h and 3p3h excitations--are taken from Ref.~\cite{Oset:1987re}, while the terms not reducible to modifications of the $\Delta$ width are taken from Ref.~\cite{Delorme:1989nh}, which exploits the calculations of Ref.~\cite{Shimizu:1980kb}. Further details are given in Appendix \ref{appendix_npnh}. We also recall that, 
for the purpose of validating the present implementation against the previously published  theoretical results, 
all the form factors entering the hadron tensor coincide with those of Ref.~\cite{Martini:2009uj}, with exactly the same parameters. For example, for the axial form factor, a standard dipole parametrization is adopted, with $M_A = \SI{1.032}{\giga\electronvolt\per\clight\squared}$ and the value\footnote{We remind the reader that the PDG 2024~\cite{ParticleDataGroup:2024cfk} value of the axial coupling constant is slightly different, being $G_A(0) = 1.2754 \pm 0.0013$. The effects of this difference with respect to our adopted value are expected to be small. They could be quantified in a future work.} of $G_A(0)=1.255$.


%% file: 3a_implementation.tex
The model described above has been implemented in the GENIE neutrino event generator using the same approach as that adopted in Refs.~\cite{Dolan:2019bxf,Dolan:2021rdd}. 
This approach relies on the fact that the differential cross section
can be expressed as the product of kinematic factors and the contraction of a generic lepton tensor with a model-specific hadron tensor. The latter encodes all of the nuclear dynamics. In this framework, the implementation of the Martini \textit{et al.} model is achieved by inserting new hadron tensor look-up tables into GENIE, as was previously done for other models in Refs.~\cite{Schwehr:2016pvn,Dolan:2019bxf,Dolan:2021rdd}. 
These tables are provided for the 1p1h and npnh channels for carbon, oxygen, and calcium targets.  The npnh contribution includes both 2p2h and 3p3h components, and separate hadron tensors are employed for each. Despite the naming, the 3p3h implementation does not produce three final-state nucleons; instead, it is treated as an additional 2p2h-like contribution and results in two nucleons in the final state. Limitations of this approach are discussed in Sec.~\ref{subsec:limitations}. The input hadron tensors are defined over an energy transfer $\omega$ varying in the range~5~$\leq \omega \leq$~\SI{995}{\mega\electronvolt} and a momentum transfer $q$ varying in the range~1~$\leq q \leq$~\SI{2000}{\mega\electronvolt\per\clight}, as in Ref.~\cite{Martini:2009uj}.
They are binned (\SI{5}{\mega\electronvolt} bins in energy transfer and \SI{20}{\mega\electronvolt\per\clight} bins in momentum transfer) and are evaluated using the same interpolation method of Refs.~\cite{Dolan:2019bxf,Dolan:2021rdd}, similar to the one described in Ref.~\cite{Schwehr:2016pvn}. 

\subsection{Validation and comparison of models}
\label{subsec:validation}
\begin{figure}[!ht]
  \centering
\includegraphics[width=\linewidth]{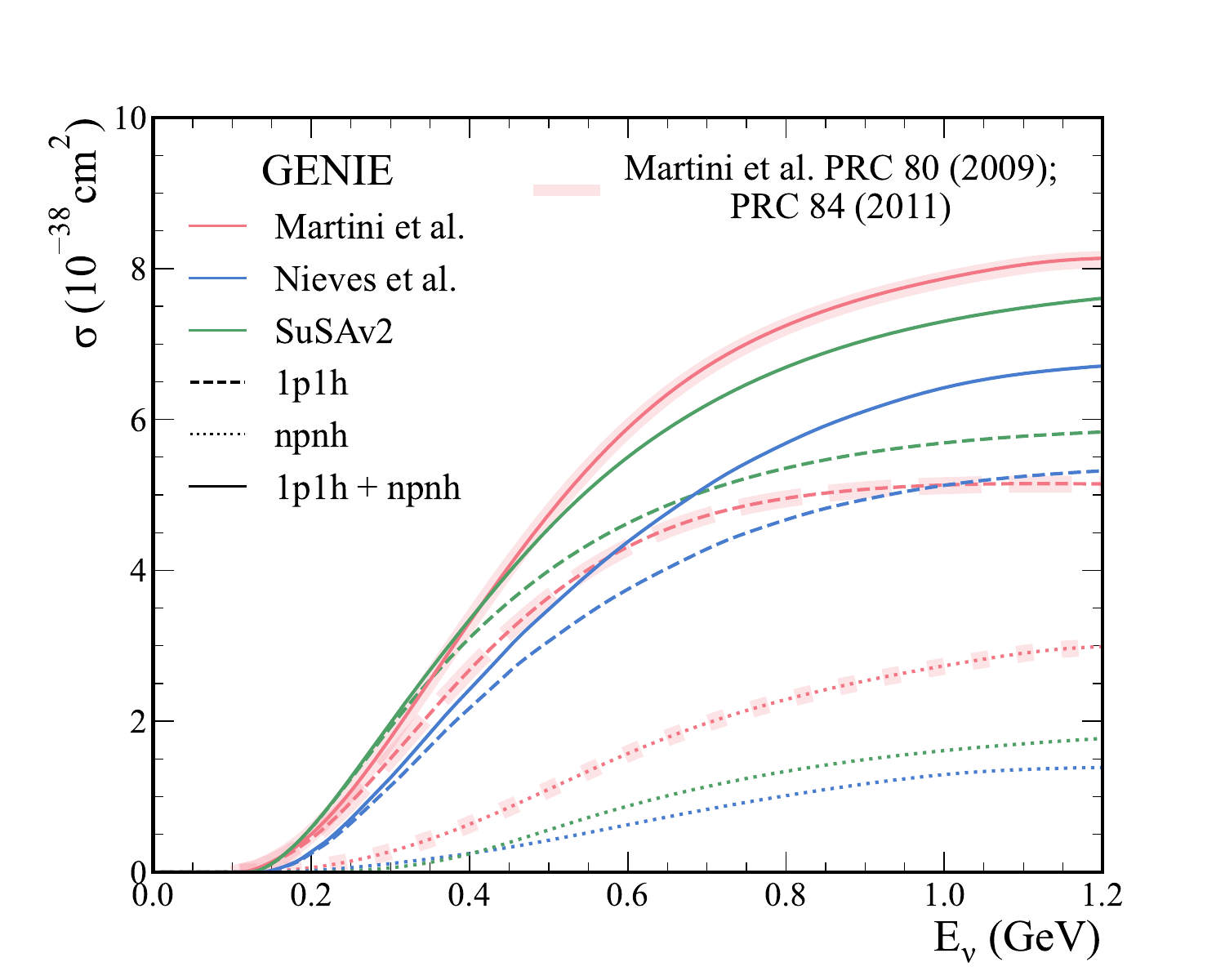}
\caption{
The total $\nu_\mu$ charged-current cross section on $^{12}$C as a function of neutrino energy predicted by the model of Martini \textit{et al.} implemented in GENIE (red). For validation purposes, the results of calculations performed by the authors in Refs.~\cite{Martini:2009uj,Martini:2011wp} are also shown (shaded band).
The different contributions to the total cross section are highlighted: 1p1h (dashed), npnh (dotted) and their sum (solid). 
The predictions of other models available in GENIE [Nieves \textit{et al.} (blue) and SuSAv2 (green)] are also shown for comparison.}
\label{fig:total_nu_allmodels}

  \centering
\includegraphics[width=\linewidth]{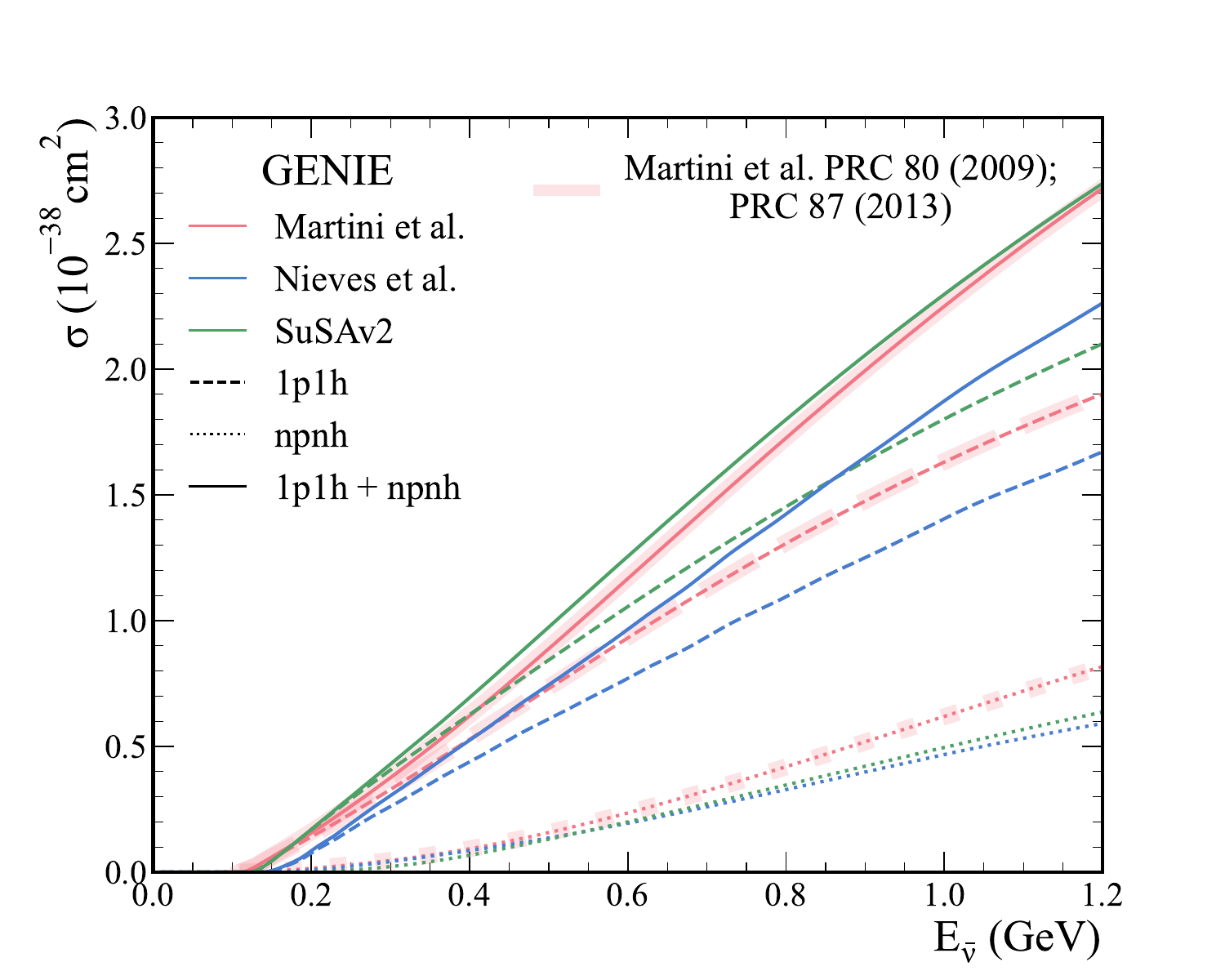}
\caption{
Same as Fig.~\ref{fig:total_nu_allmodels}, but for antineutrinos. In this case, the calculations performed by the authors are the ones of Refs.~\cite{Martini:2009uj,Martini:2013sha}.}
\label{fig:total_anu_allmodels}
\end{figure}

For validation purposes, the model predictions obtained via GENIE have been compared with the original calculations performed by the authors of the model~\cite{Martini:2009uj}, including the relativistic corrections for the quasielastic, published in Ref.~\cite{Martini:2011wp} for neutrinos and in Ref.~\cite{Martini:2013sha} for antineutrinos. The corresponding charged-current cross sections on $^{12}$C as a function of (anti)neutrino energy are presented for $\nu_\mu$ and $\bar{\nu}_\mu$ in Figs.~\ref{fig:total_nu_allmodels} and \ref{fig:total_anu_allmodels}, respectively (and in Figs.~\ref{fig:total_nu_alltargets} and \ref{fig:total_anu_alltargets} of Appendix~\ref{app} for $^{16}$O and $^{40}$Ca). These plots demonstrate perfect agreement between the original calculations (shaded bands) and the GENIE predictions for the different contributions to the cross section: ``1p1h'' (dashed), ``npnh'' (dotted) and their sum (solid).
The predictions of other 1p1h and 2p2h models available in GENIE (Nieves \textit{et al.} and SuSAv2) are also shown for reference. The corresponding GENIE tunes used are \texttt{G18\_10a\_00\_000} for the Nieves \textit{et al.} model and \texttt{G21\_11a\_00\_000} for SuSAv2. As already illustrated in other articles (see, for example, Fig.~9 of Ref.~\cite{T2K:2023qjb}), the npnh (as well as the 1p1h) contributions estimated by these three models are different, and their relative importance with respect to 1p1h varies between neutrinos and antineutrinos, as discussed in Refs.~\cite{Martini:2010ex,Katori:2016yel}. It is also important to note that a one-to-one correspondence between different exclusive channel contributions can be misleading. For example the nucleon-nucleon short-range correlations are part of npnh channel in the Martini \textit{et al.} model, whereas they are included in the 1p1h component for SuSAv2. As can be seen in Figs.~\ref{fig:total_nu_allmodels} and \ref{fig:total_anu_allmodels}, the sum of these two channels is similar for Martini \textit{et al.} and SuSAv2, especially in the antineutrino case.

\begin{figure}[!ht]
  \centering
\includegraphics[width=\linewidth]{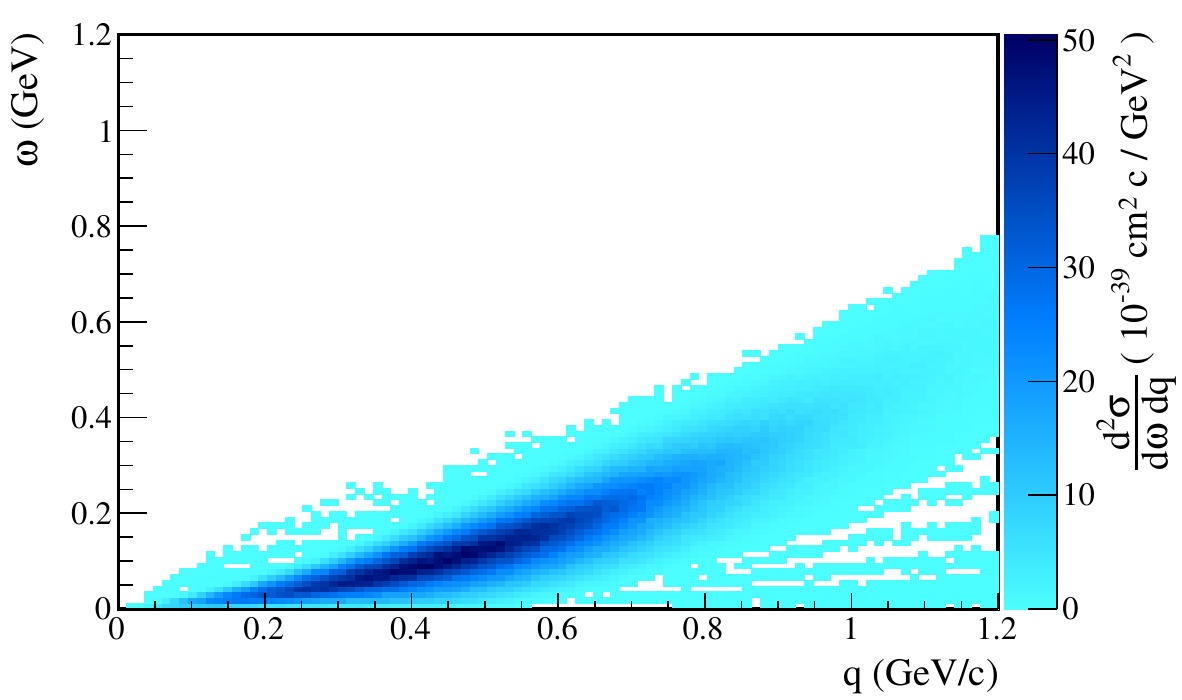}
  \caption{The T2K flux-integrated $\nu_\mu$ double differential cross section as a function of $q$ and $\omega$ for the 1p1h excitation channel, obtained from the GENIE implementation of the calculations in Ref.~\cite{Martini:2011wp}.}
\label{fig:q0q3CCQE}

  \centering
\includegraphics[width=\linewidth]{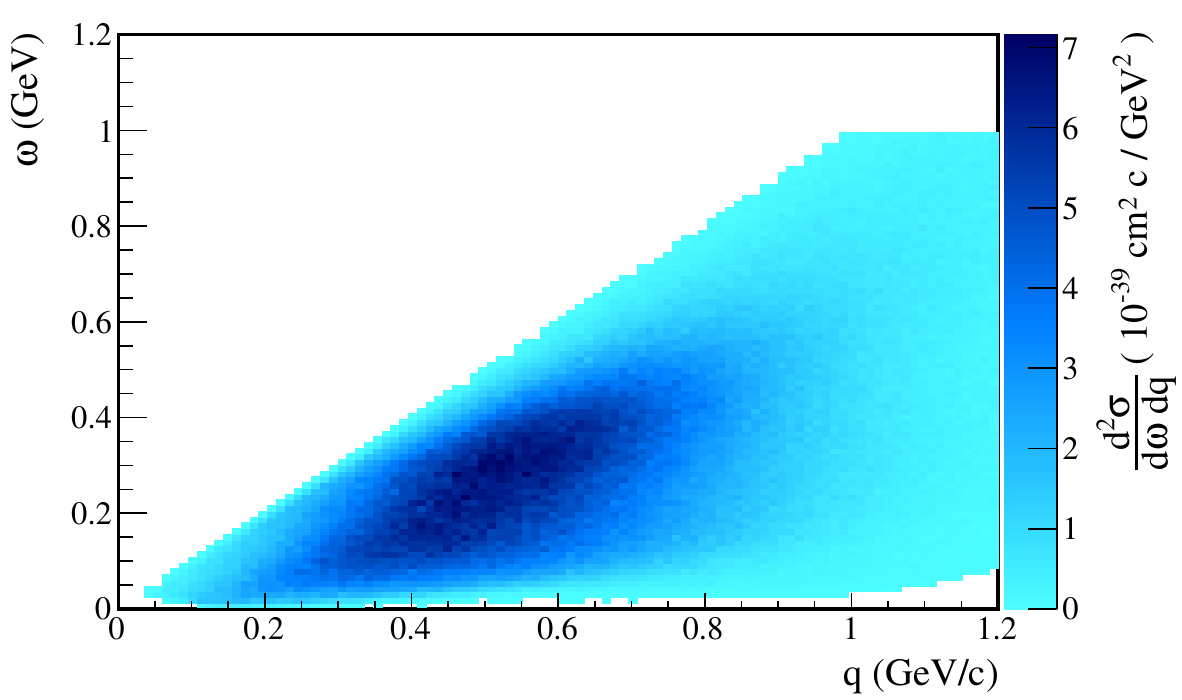}
  \caption{The T2K flux-integrated $\nu_\mu$ double differential cross section as a function of $q$ and $\omega$ for the npnh excitation channel, obtained from the GENIE implementation of the calculations in Ref.~\cite{Martini:2009uj}.}
\label{fig:q0q3CCMEC}
\end{figure}

For the sake of comparison with other models, we show in Figs.~\ref{fig:q0q3CCQE} and \ref{fig:q0q3CCMEC} the T2K flux-integrated $\nu_\mu$ double differential cross section as a function of $q$ and $\omega$ for the 1p1h and npnh channels, corresponding to Fig.~4 of \cite{Dolan:2021rdd} and Fig.~2 of \cite{Dolan:2019bxf}, respectively. In Fig.~\ref{fig:q0q3CCQE}, one can easily recognize the quasielastic line
 \begin{equation}
    \label{eq_omega_qe}
       \omega=\frac{q^2-\omega^2}{2M_N}=\sqrt{q^2+M_N^2}-M_N 
\end{equation}
     where $M_N$ denotes the nucleon mass. The spread around this curve is due to Fermi motion\footnote{\label{fn:scatterplot}The fact that the quasielastic region is not strictly limited by the two quasielastic bands
$\omega_{\pm}=\sqrt{q^2\pm2qk_F+M_N^2}-M_N$, 
typical of a global relativistic Fermi gas-based model with Fermi momentum $k_F$, is due to the local density approximation and to the presence of Bessel functions in the nuclear response definitions (see Appendix 2 of Ref.~\cite{Martini:2009uj}), which result in very small fluctuations beyond the quasielastic region.}. From Fig.~\ref{fig:q0q3CCMEC} one can observe that the multinucleon response region covers the whole available $(q,\omega)$ phase space from the input tensors. Furthermore, two distinct branches can be identified: one below the quasielastic region--absent in the SuSAv2 case, see Fig.~2 of Ref.~\cite{Dolan:2019bxf}--arising from nucleon-nucleon correlations, and a second one above the quasielastic region, extending from the dip to the $\Delta$ region, due to $\Delta$-MEC contributions\footnote{These two branches are more distinguishable looking directly to the phase space of the nuclear response function, rather than to the flux-integrated cross section. See for example Fig.~7 of Ref.~\cite{Katori:2016yel}.}. Overall, these two branches are less clearly separated than in the Nieves \textit{et al.} model (see again Fig.~2 of Ref.~\cite{Dolan:2019bxf}); in the Martini \textit{et al.} case, the cross section is concentrated between the two contributions, which add coherently together with their significant interference. To summarize, the phase-space behavior of the multinucleon cross section in the Martini \textit{et al.} model differs from those of Nieves \textit{et al.} and SuSAv2, though it shares features with both: the presence of two branches, as in Nieves \textit{et al.}, and the accumulation of cross-section strength in the dip region, as in SuSAv2.

\begin{figure*}[!ht]
  \centering
  \includegraphics[width=0.49\linewidth]{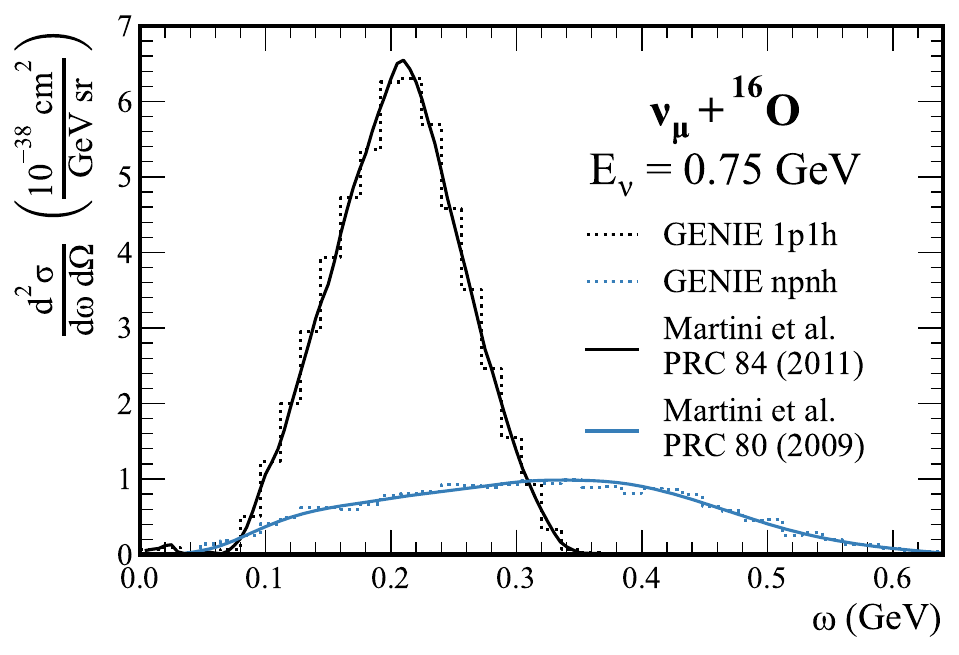}
  \includegraphics[width=0.49\linewidth]{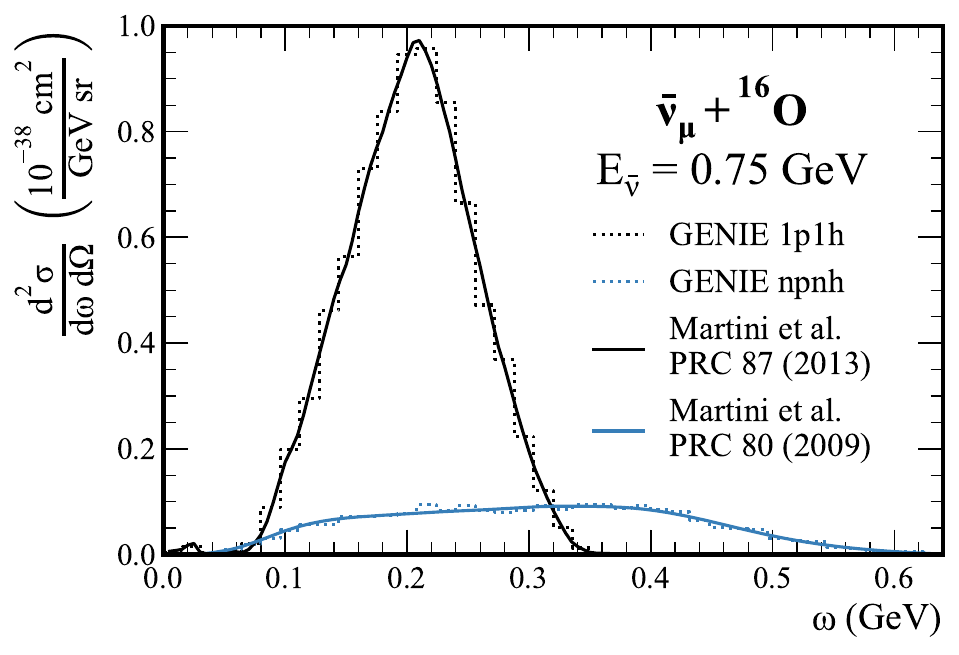}
  \caption{Double differential cross sections 
  $\frac{d^2\sigma}{d\omega\, d\Omega}$
  for 1p1h (black) and npnh (blue) processes in muon neutrino (left) and muon antineutrino (right) interactions with oxygen, at a scattering angle of $60^\circ$ and incident (anti)neutrino energy of \SI{0.75}{\giga\electronvolt}. The cross section is plotted as a function of the energy transfer $\omega$ to the nucleus. Continuous lines: theoretical calculations of Refs.~\cite{Martini:2009uj,Martini:2011wp,Martini:2013sha}; histograms: the corresponding GENIE implementation.}
  \label{fig:ddxsec_E0.75_O16}
\end{figure*}

\begin{figure*}[!ht]
  \centering
\includegraphics[width=\linewidth]{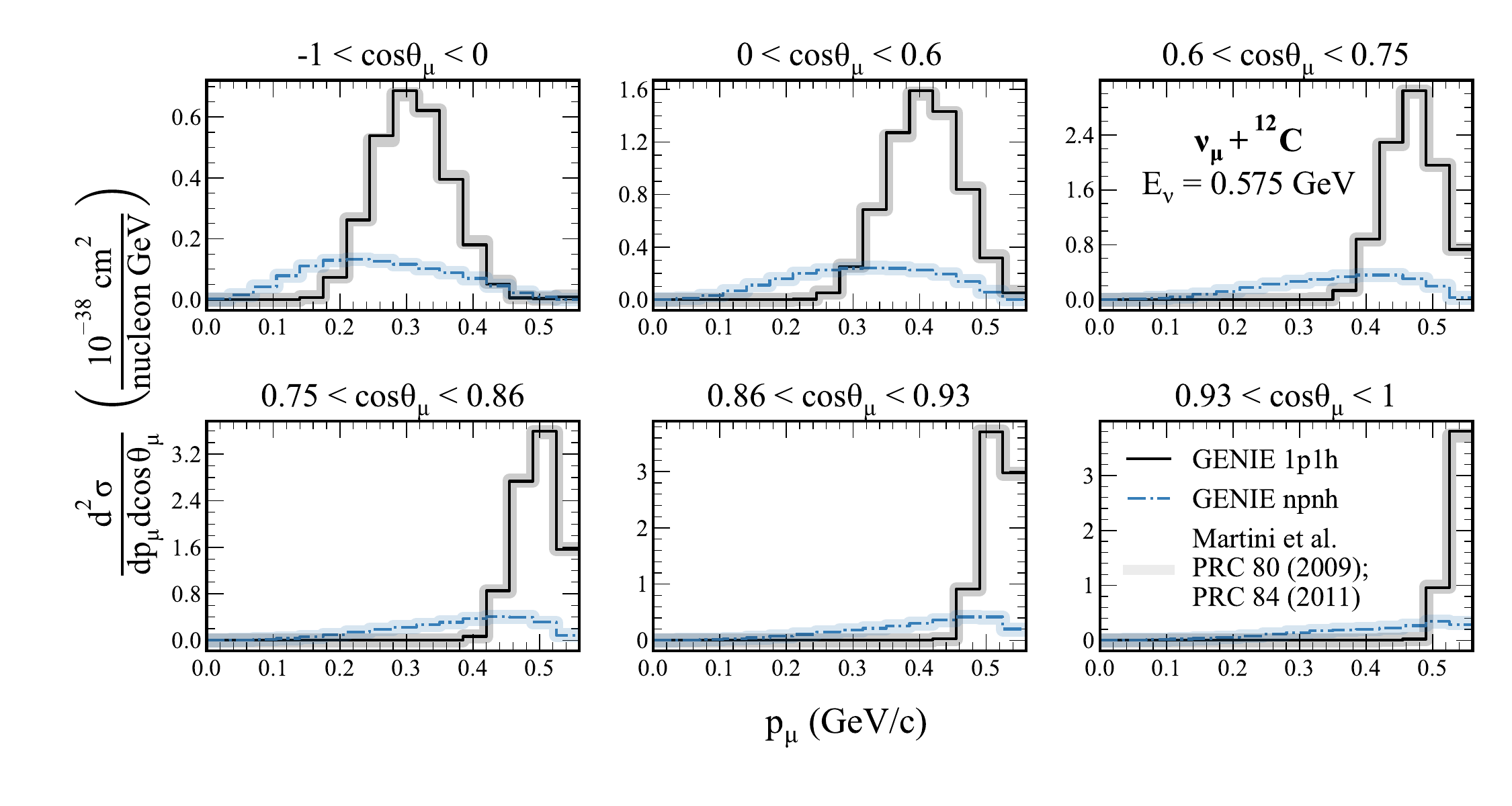}
\caption{Double differential cross sections $\frac{d^2 \sigma}{dp_\mu \; d\!\cos\!\theta_\mu}$ per nucleon, as a function of outgoing muon momentum, for 1p1h and npnh processes in muon neutrino interactions with carbon, at an incident neutrino energy of \SI{0.575}{\giga\electronvolt} and for different scattering angle regions. The shaded bands correspond to theoretical calculations of Refs.~\cite{Martini:2009uj,Martini:2011wp}; black continuous (1p1h) and blue dash-dotted (npnh) lines show the corresponding GENIE implementation.}
\label{fig:ddE0.575_C12}
\end{figure*}

As a validation of the implementation, and to facilitate comparisons with other models, in Fig.~\ref{fig:ddxsec_E0.75_O16} we present the double differential cross sections\footnote{Here, $d\Omega$ denotes the solid angle element. Assuming azimuthal symmetry, the double differential cross section is related by $\frac{d^2\sigma}{d\omega\, d\Omega} = \frac{1}{2\pi} \frac{d^2\sigma}{d\omega\, d\!\cos\!\theta}$.} $\frac{d^2\sigma}{d\omega\, d\Omega}$ for both neutrino and antineutrino interactions on oxygen, at fixed incident energy and lepton scattering angle,\footnote{GENIE does not allow the user to explicitly fix the scattering angle. Events with $0.495 \leq \cos\theta_{\mu} \leq 0.505$ were selected to approximate a scattering angle of $60^\circ$.} as a function of the energy transfer $\omega$. 
This represents the most stringent validation test, as the cross sections are not integrated over any variable, in contrast to the total or flux-integrated differential cross sections discussed earlier. A perfect agreement is observed between the theoretical calculations (solid lines) and the GENIE implementation (histograms) for both 1p1h and npnh channels, and for neutrino as well as antineutrino interactions.\footnote{In the 1p1h case, the small fluctuations at low $\omega$ discussed in footnote \ref{fn:scatterplot} can also be seen.}

Figure~\ref{fig:ddxsec_E0.75_O16} can be directly compared with Fig.~9 of Ref.~\cite{Sobczyk:2024ecl}, the recent reexamination by Sobczyk and Nieves of the multinucleon cross-section model originally presented in Ref.~\cite{Nieves:2011pp}. In Ref.~\cite{Sobczyk:2024ecl}, the revised npnh single differential cross section $\frac{d\sigma}{d\omega}$ at fixed neutrino energy, as well as the MiniBooNE $\nu_\mu$ flux-integrated double differential cross section $\frac{d^2\sigma}{dT_\mu\,d\!\cos\!{\theta_\mu}}$ on carbon, are compared with the corresponding results of Martini \textit{et al.}~\cite{Martini:2009uj,Martini:2011wp}. The updated Valencia predictions are shown to be closer to the Lyon model, thus ending a long-standing discrepancy between the two.

We now extend the comparison to the double differential cross section at fixed neutrino energy. Before focusing on the npnh channel, we note that the 1p1h distribution in Fig.~\ref{fig:ddxsec_E0.75_O16} exhibits a shape similar to that in Fig.~9 of Ref.~\cite{Sobczyk:2024ecl}, as expected, since both are based on RPA calculations on top of a local relativistic Fermi gas. However, the distribution in Fig.~\ref{fig:ddxsec_E0.75_O16} is slightly more peaked and shifted to lower $\omega$ by approximately \numrange{10}{20}\,\si{\mega\electronvolt}. This is probably due to the absence of removal energy in the calculations of Martini \textit{et al.}, unlike the case of Nieves \textit{et al.} These small differences are often washed out in flux-integrated double differential distributions (see, e.g., Fig.~10 of Ref.~\cite{Sobczyk:2024ecl}).

Turning to the npnh channel, for neutrinos, the evaluation of Ref.~\cite{Sobczyk:2024ecl} aligns more closely with the results of Ref.~\cite{Martini:2009uj} than with the earlier ones of Ref.~\cite{Nieves:2011pp}. However, the double-bump structure shown in Fig.~9 of Ref.~\cite{Sobczyk:2024ecl}, attributed to nucleon-nucleon correlations and MEC contributions, is not visible in Fig.~\ref{fig:ddxsec_E0.75_O16}.
For antineutrinos, 
the npnh contribution in Fig.~\ref{fig:ddxsec_E0.75_O16} is smaller than that in Fig.~9 of Ref.~\cite{Sobczyk:2024ecl}; nonetheless, the total cross section as a function of antineutrino energy in Fig.~8 of Ref.~\cite{Sobczyk:2024ecl} appears to be closer to the results of Ref.~\cite{Martini:2009uj} than to the previous Valencia model predictions shown in Fig.~\ref{fig:total_anu_allmodels}.

To summarize, although the revised npnh model of Valencia reported in Ref.~\cite{Sobczyk:2024ecl} generally yields results more compatible with those of Martini \textit{et al.}, some discrepancies remain. These differences could be precisely quantified once the revised approach of Ref.~\cite{Sobczyk:2024ecl} is implemented in GENIE.

Figure~\ref{fig:ddxsec_E0.75_O16} provides a specific example of a double differential cross section at fixed kinematics. To further illustrate the perfect correspondence between the theoretical calculations and the GENIE implementation in other configurations, we show in Fig.~\ref{fig:ddE0.575_C12} the results for carbon at a different incident neutrino energy and for lepton scattering angle bins corresponding to those used in the T2K measurement \cite{T2K:2020jav} discussed in the next section.

%% file: 3b_limitations.tex
\subsection{Limitations of the model implementation}
\label{subsec:limitations}

While the implementation described here represents 
an inclusion of the 1p1h and npnh channels of the Martini \textit{et al.} model within the GENIE event generator for a set of defined nuclear targets for charged-current neutrino and antineutrino scattering, there are limitations that are important to note. Some of these are related to approximations inherent in producing a fully exclusive simulations from an inclusive input model, which are common to many model implementations in neutrino interaction generators. Others are due to missing predictive power within the Martini \textit{et al.} model itself. Some of these limitations can be mitigated with relatively modest extensions of the implementation.

First, it is important to note that the hadron tensors for both 1p1h and npnh interactions currently only go up to \SI{995}{\mega\electronvolt} in energy transfer and \SI{2000}{\mega\electronvolt\per\clight} in momentum transfer. This is due to a choice made in the original implementation of the model~\cite{Martini:2009uj}.
In the current GENIE implementation, outside of this range, the differential cross section vanishes (like for the other models following the hadron tensor implementation method). Further tractable but nontrivial model development work would be required to overcome this limitation. In principle, the implementation can be artificially expanded by interpolating (``blending'') the cross sections predicted by the Martini \textit{et al.} model with those from another inclusive model, such as SuSAv2 (as already done within the CRPA model implementation scheme in GENIE~\cite{Dolan:2021rdd}). For the purposes of this implementation, as well as the comparisons with flux-averaged measurements we consider in this work, the current implementation is sufficient. We remind the reader also that, even if the tensors start from $\omega=~$\SI{5}{\mega\electronvolt}, 
giant resonance peaks that may appear at low transferred momenta for $\omega \lessapprox$ \SI{50}{\mega\electronvolt} are not included in the formalism of Martini \textit{et al.} at difference of the CRPA ~\cite{Pandey:2014tza,Dolan:2021rdd} approach, particularly suited for this low-energy low-momentum region. However, in this region, the RPA Martini \textit{et al.} calculations (which do not include giant resonance peaks) may represent a reasonable average of the CRPA results~\cite{Pandey:2014tza}, as shown in Fig.~2 of Ref.~\cite{Martini:2016eec} and discussed in that article.

Importantly, the implementation scheme described in this document relies on a factorization scheme to generate hadron kinematics from an inclusive input model, like many implementations in neutrino interaction generators. The scheme followed is identical to that of Refs.~\cite{Dolan:2019bxf, Dolan:2021rdd} and is discussed further in Ref.~\cite{Nikolakopoulos:2023pdw}. This has only very small effects on the cross-section predictions shown within this paper since they have either no or only small dependence on the hadron kinematics predicted by GENIE. 

We additionally note that, although the implementation features a 3p3h contribution to the npnh channel, this has been added as an additional ``effective'' 2p2h channel. This means that only two nucleons are produced at the vertex for a 3p3h interaction, relying solely on the existing GENIE hadron kinematics generator. This could be mitigated in a future extension to this implementation to produce a third final-state nucleon (via a new interaction channel within GENIE)\footnote{We note that separate 2p2h and 3p3h hadron tensors are available if the user prefers to only simulate one of the two npnh contributions, or a specific sub-contribution within the 2p2h channel as described in the final paragraph of Sec.~\ref{sec:theo}. This option can be configured by editing the \texttt{config/MartiniMECHadronTensorModel.xml} configuration file.}. This was not implemented at the time of writing this article as the current implementation uses the same routine to process the combined npnh (i.e. 2p2h+3p3h) tensors. Separating the two types of interactions would have required creating a new routine for the 3p3h channel. Since the current model predictions are inclusive, we leave this extension for future work. Again, predictions in this article are either entirely unaffected or not very sensitive to alterations to the treatment of the hadronic system.

The current implementation relies on hadron tensors computed for three isoscalar nuclei--$^{12}$C, $^{16}$O and $^{40}$Ca. For nuclear targets other than those for which the tensors have been calculated, the total 1p1h cross section is scaled according to the difference in the number of target nucleons (i.e. depending on whether the incoming probe is a neutrino or an antineutrino). For example, for an $^{40}$Ar target, the total 1p1h cross section on $^{40}$Ca is scaled by a factor of 22/20 (18/20) for an incoming neutrino (antineutrino) probe. 
For npnh interactions, the total cross section is scaled with the mass number, also following the scheme in Ref.~\cite{Dolan:2019bxf}, implying no scaling between the $^{40}$Ca and $^{40}$Ar npnh cross sections. Note that a small additional scaling is applied, following the same scheme as for the SuSAv2 model~\cite{Dolan:2019bxf} based on superscaling arguments. This  is the (square of the) ratio of the Fermi momenta of the reference and target nuclei for 1p1h (npnh) interactions.\footnote{The Fermi momentum used for this calculation are taken from Ref.~\cite{Maieron:2001it}. It is assumed to be the same for protons and neutrons. For nuclei not in the table the Fermi momentum of the nucleus with the closest proton number is used.} While not perfect, this overall scaling is expected to be a reasonable approximation away from the low-energy transfer region ($\lessapprox$100 MeV). Scaling to heavier nuclear targets occasionally employed in neutrino scattering measurements, such as iron or lead, could be improved with a future addition of a hadron tensor table for a heavier target.

More generally, the Martini \textit{et al.} model internally does not differentiate between protons and neutrons when calculating cross sections, meaning that the impact of nonisoscalarity of target nuclei is not accounted for. Mitigating this limitation can be a challenging task. A first step could be assigning different Fermi momenta ($k_F$) to neutrons and protons within the model. However, this naive scaling does not take into account the RPA modification for asymmetric nuclei \cite{Alberico:1989zz}, a challenge that, to our knowledge, is shared with the other RPA-based models at present, theoretical calculations of asymmetric relativistic Fermi gas for quasielastic lepton-nucleus scattering being only relatively recent \cite{Barbaro:2018kxa}. It is also important to stress that the first microscopic calculation of two-particle-two-hole meson-exchange currents in $^{40}$Ar discussing also asymmetric scaling properties for neutrino and electron scattering appeared only very recently \cite{Martinez-Consentino:2025ebw}.



Finally, the Martini \textit{et al.} model, and therefore the hadron tensors implemented in GENIE, do not take into account the impact of the nuclear removal energy. The consequence of this is that the lepton kinematics, in both GENIE and the Martini \textit{et al.} model, are calculated assuming an interaction with unbound (but still moving) nucleons. On the other hand, GENIE does consider the removal energy when calculating hadron kinematics via the aforementioned factorization scheme. While in this paper we do not attempt to modify the original inclusive predictions of the Martini \textit{et al.} model, a possible mitigation for this limitation could be an extension of this implementation to add an effective modification to the energy transfer read from the hadron tensors to simulate the impact of removal energy. This strategy can be trivially adopted using routines already available within the GENIE generator.

Overall, the model implementation is a faithful reproduction of the Martini \textit{et al.} model for outgoing lepton kinematics for carbon, oxygen or calcium nuclei. Scaling to different nuclei of similar atomic number is broadly reliable. The hadron kinematics predicted by GENIE from this model implementation should be treated with cautious skepticism (as is the case also for other 1p1h and 2p2h GENIE model implementations).

%% file: 4_datacomparison.tex

As discussed in Sec.~\ref{sec:theo}, the theoretical calculations by Martini \textit{et al.} have already been compared in the past with several cross-section measurements. However, previous comparisons were performed in the absence of a unified generator implementation, so the treatment of effects such as final-state interactions (FSIs) was not applied in a consistent way to all interaction channels. Here we present a few additional comparisons to illustrate the potential of the present GENIE implementation to make cross-section predictions for different neutrino beams, various nuclear targets and diverse final-state topologies, using a unified generator framework as enabled by this implementation. 

Throughout this section, we compare to two experimental measurements from the T2K and MicroBooNE Collaborations, both of which report the cross section for the so-called CC$0\pi$ topology. This topology contains contributions from the 1p1h and npnh channels, which are simulated with the Martini \textit{et al.}~model implementation described in this article. There is also a subleading contribution from interactions where a pion was produced (usually via a resonant interaction) and then absorbed via FSI. To model this contribution, we use the standard GENIE prescription (Ref.~\cite{GENIE:2021npt}) which employs the Berger-Sehgal model (Ref.~\cite{Berger:2007rq}) with updated form factors from Graczyk and Sobczyk (Ref.~\cite{Graczyk:2007bc}) and includes  a variety of resonances up to invariant hadronic mass $W = 1.7$~GeV, but neglecting interferences between them. Final-state interactions are simulated via the GENIE intranuclear cascade using the \texttt{hA2018Intranuke} routine (for details, see Ref.~\cite{Dytman:2021ohr}). FSIs will impact the predictions in several ways. Although the FSI model in GENIE does not modify the total inclusive cross section, it will affect the hadrons that are used in defining the signal topologies. For both T2K and MicroBooNE, FSI impacts the samples mainly via pion absorption, which contributes to the measured signal sample. In addition to this, the MicroBooNE samples require at least one proton with momentum above \SI{300}{\mega\electronvolt\per\clight}, and the final-state proton kinematics that determine the sample composition will also be impacted by the choice of the FSI model (its effect is notably to decelerate protons and thus reduce the total $\rm CC1p0\pi$ cross section). It is also worth noting that the MicroBooNE $\rm CC1p0\pi$ samples also allow final-state charged pions with momenta below \SI{70}{\mega\electronvolt\per\clight}, which can also be affected by the choice of resonant pion production and FSI models. Finally, we note that FSI can also impact both T2K and MicroBooNE samples via pion production from nucleon interactions with the residual nucleus, but the contribution of these processes in this energy range is negligible for the purposes of this discussion.

\begin{figure*}[htb]
  \centering
  \begin{subfigure}{\textwidth}
    \centering
    \includegraphics[width=\linewidth]{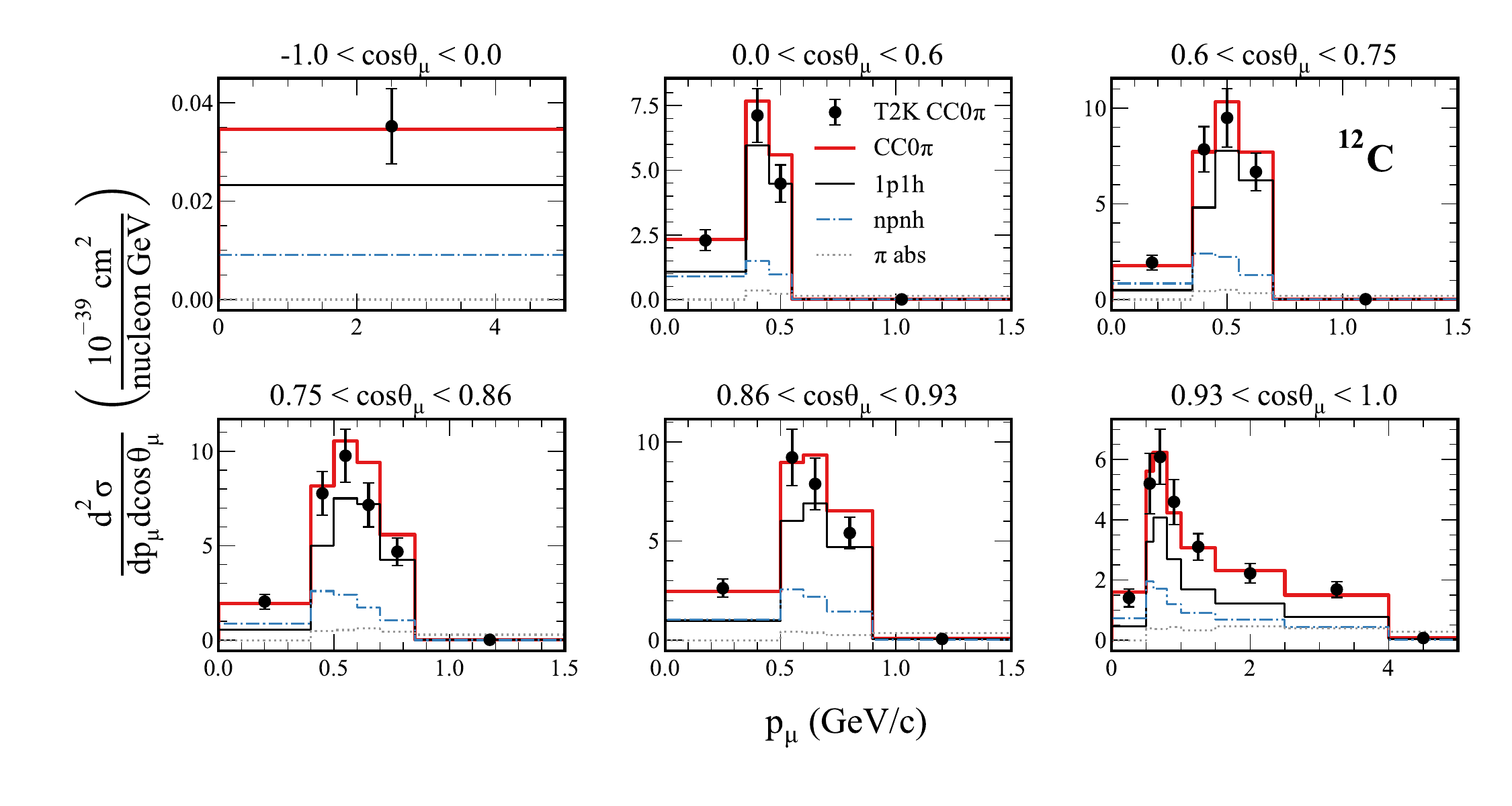}
    \caption{Carbon target}
    \label{fig:T2K_C12}
  \end{subfigure}
  \hfill
  \begin{subfigure}{\textwidth}
    \centering
    \includegraphics[width=\linewidth]{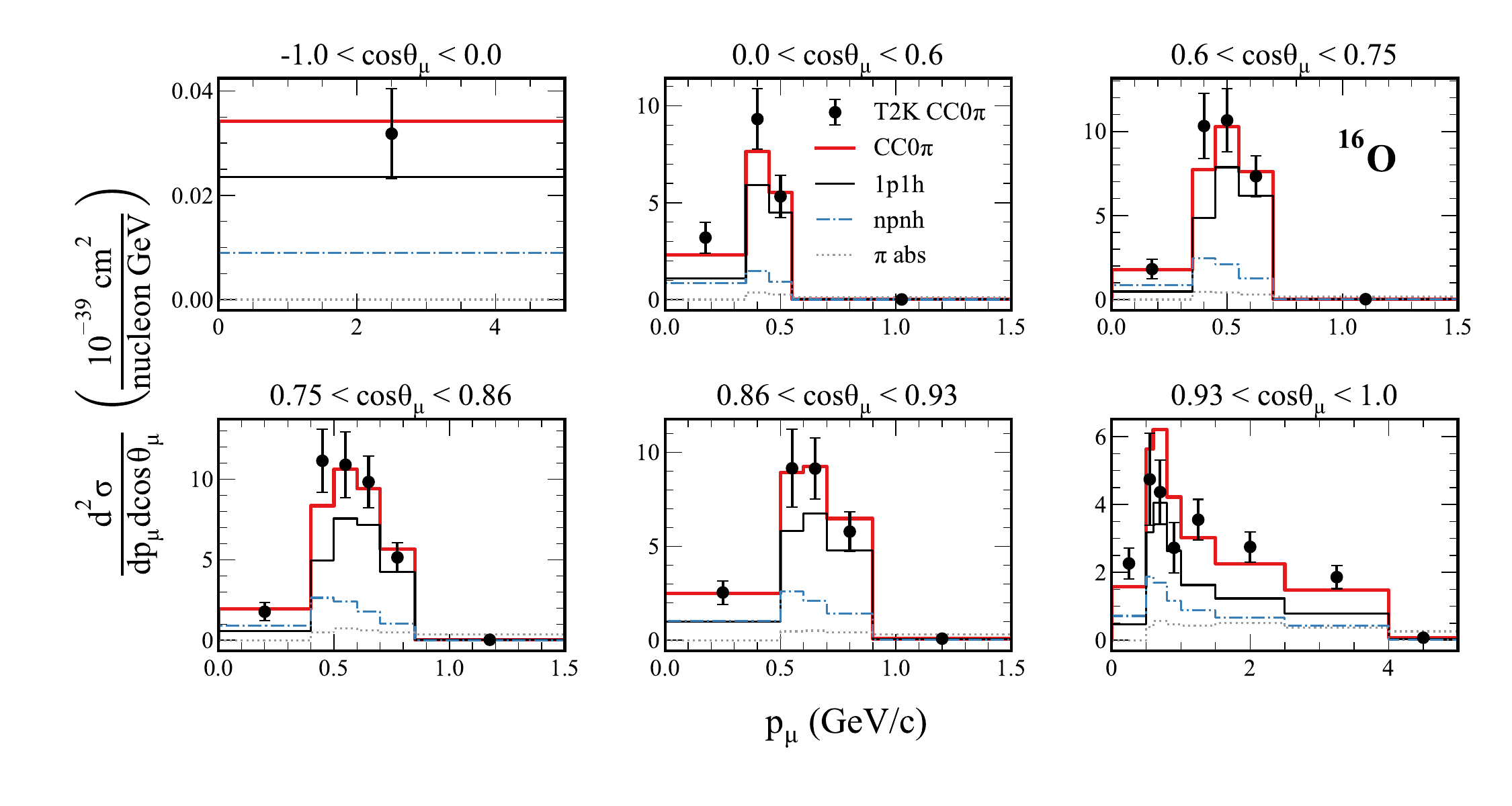}
    \caption{Oxygen target}
    \label{fig:T2K_O16}
  \end{subfigure}
  \caption{T2K flux-integrated $\nu_\mu$ charged-current double differential cross sections $\frac{d^2 \sigma}{d p_\mu \; d\!\cos \!\theta_\mu}$ without pions in the final state, measured per nucleon on (a) carbon and (b) oxygen targets~\cite{T2K:2020jav}. The measurements (points with error bars) are compared with the GENIE implementation of the Martini \textit{et al.} model. The contributions shown are: 1p1h (thin solid black), npnh (dash-dotted), the pion absorption ``$\pi$ abs'' (dotted gray), and the total CC0$\pi$ (thick solid red), which is the sum of the previous ones. For all histograms, the last data point (which covers the energy range up to 30~GeV) is shifted to the middle of the presented energy bin. The total $\chi^2$ for the joint carbon and oxygen measurement is $\chi^2 = 76.6$ with 58 degrees of freedom.}
  \label{fig:T2K_C12_O16}
\end{figure*}

We begin by showing the T2K flux-integrated $\nu_\mu$ charged-current double differential cross sections $\frac{d^2 \sigma}{dp_\mu \; d\!\cos\!\theta_\mu}$ without pions in the final state, i.e. the ``CC0$\pi$'' cross section, on carbon (Fig.~\ref{fig:T2K_C12}) and oxygen (Fig.~\ref{fig:T2K_O16}). The experimental T2K results are the regularized cross sections obtained via simultaneous analysis of the carbon and oxygen target samples, see Ref.~\cite{T2K:2020jav} (and Fig.~9 in this publication) for more details. 
To process the simulated events, which include pion absorption contribution as implemented in GENIE, we use the NUISANCE framework~\cite{Stowell:2016jfr} which  facilitates direct measurement-model comparisons. 

The Martini~\textit{et al.} model yields a reasonable agreement with the T2K measurement. The total $\chi^2$ for both carbon and oxygen distributions is $\chi^2=76.6$ for 58 degrees of freedom,\footnote{A similar comparison with the so-called unregularized cross sections~\cite{T2K:2020jav} yields $\chi^2=73.4$ for 58 degrees of freedom.} which corresponds to a \textit{p} value of 0.05, indicating satisfactory agreement. This result can be compared with the $\chi^2$ values presented in Table~V of Ref.~\cite{T2K:2020jav} which presents the $\chi^2$ for nine different versions/tunes of the neutrino event generators.\footnote{Note that some of these models have been updated since the measurement publication.} The $\chi^2$ obtained with the Martini~\textit{et al.} model is lower than that of five of these models and higher than that of four. It is useful to note that Table~V of Ref.~\cite{T2K:2020jav} emphasizes that the majority of the discrepancy between the models and the measurement is driven by the most forward $\text{cos}\theta_\mu$ bin, which is dominated by low-energy transfer interactions. This is the region where suppression of the cross section from long-range correlations via RPA is most pronounced. 

The Martini~\textit{et al.}~model generally performs better than both the NEUT and NuWro spectral function models. This is expected, as these model implementations do not account for suppression of the cross section via RPA effects. The Martini~\textit{et al.} model also performs better than the GENIE SuSAv2 implementation and the RMF 1p1h+SuSAv2 prediction. In the first case, the SuSAv2 1p1h model implementation in GENIE does not include an explicit RPA suppression (but does include an effective suppression of the cross section at low-energy transfer due to FSI; for more details, see~\cite{Dolan:2019bxf}), which results in an overprediction of the measurement in the most forward $\text{cos}$$\theta_\mu$ bins. In the case of the RMF model, the effect of FSI is included in the distortion of the final wave function from first principles, which causes a larger reduction in the cross section and thus a better agreement with the measurement. The Martini~\textit{et al.} model performs better than both of these models, indicating the probable need for an even larger suppression due to RPA effects. Finally, our implementation yields a lower $\chi^2$ than the GiBUU model tested in Ref.~\cite{T2K:2020jav}. This may be due to the fact that the LFG model used in GiBUU~\cite{Buss:2011mx} does not apply any RPA effects.

\begin{figure}[!ht]
  \centering
\includegraphics[width=0.99\linewidth]{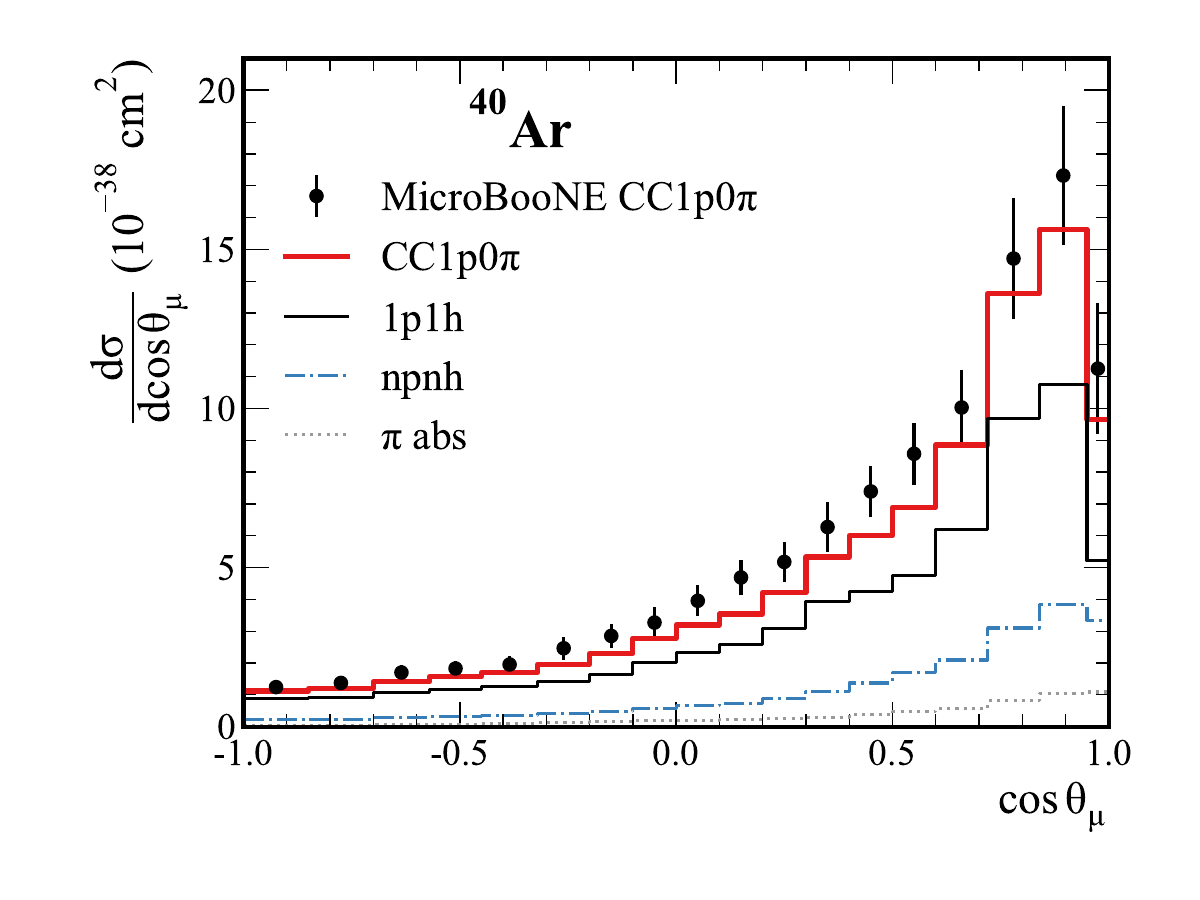}
  \caption{MicroBooNE measurements of the flux-integrated $\nu_\mu$ charged-current single differential cross sections $\frac{d\sigma}{d\!\cos\!\theta_\mu}$ without pions and one proton in the final state \cite{MicroBooNE:2023cmw}, the so-called CC1p0$\pi$ cross section, on argon compared with the GENIE implementation of the Martini \textit{et al.} model. Different contributions are shown: 1p1h (thin solid black), npnh (dash-dotted blue), the pion absorption $\pi$ abs (dotted gray), and the total CC1p0$\pi$ (thick solid red), which is the sum of the previous ones. A total $\chi^2$ of 6.4 is obtained for 18 degrees of freedom.}
\label{fig:microboone_nu}
\end{figure}

Conversely, the Martini~\textit{et al.} model yields a higher $\chi^2$ than the other GENIE and NEUT~\cite{Hayato:2021heg} LFG+RPA-based models. In both cases, these implementations are based on the model of Nieves \textit{et al.}~\cite{Nieves:2011pp} and, as discussed in Sec.~\ref{subsec:validation}, have similar predictions for the inclusive cross section on carbon to those of the Martini~\textit{et al.} model (see Figs.~\ref{fig:total_nu_allmodels} and~9 of Ref.~\cite{Sobczyk:2024ecl}). A significant portion of the difference in the $\chi^2$ between the LFG-based models and the Martini~\textit{et al.} model is therefore likely due to the treatment of 2p2h interactions. Whereas the updated treatment of multinucleon interactions discussed in Ref.~\cite{Sobczyk:2024ecl} reduces the disagreement between the Martini~\textit{et al.} and the Nieves~\textit{et al.} models, the treatment used in the comparisons of Ref.~\cite{T2K:2020jav} predicted a 2p2h cross section for the Nieves \textit{et al.} model which is significantly lower than that of the Martini~\textit{et al.} (as shown in Fig.~\ref{fig:total_nu_allmodels}).  It is additionally worth noting that the quoted $\chi^2$ values are for the \textit{combined} carbon and oxygen measurement. A closer inspection of the individual carbon and oxygen inclusive predictions in the most forward $\text{cos}\theta_{\mu}$ bin reveals that the Nieves \textit{et al.} and Martini~\textit{et al.} predictions are extremely close for carbon, but they differ slightly for oxygen. 
This implies that some missing ingredients in the detailed differences between the cross sections on carbon and oxygen affecting the forward $\text{cos}\theta_\mu$ region (for example concerning the nuclear removal energy) may be responsible for the higher $\chi^2$.

Another example of model-measurement comparison is shown in Fig.~\ref{fig:microboone_nu}, which presents the MicroBooNE flux-integrated $\nu_\mu$ charged-current single differential cross sections $\frac{d\sigma}{d\!\cos\!\theta_\mu}$ without pions and with one proton in the final state--the so called CC1p0$\pi$ cross section--on argon. Compared to the previous T2K case, this measurement involves a different neutrino flux, a different nuclear target, and a different final-state topology. The measurement is that published by MicroBooNE in Ref.~\cite{MicroBooNE:2023cmw}, where the signal includes all $\nu_\mu$-Ar scattering events with a final-state muon of momentum $0.1< p_\mu< 1.2$ \SI{}{\giga\electronvolt\per\clight} and exactly one proton with $0.3 < p_p < 1$ \SI{}{\giga\electronvolt\per\clight}. Regarding the modeling, we recall that the implemented microscopic calculations are those of $^{40}$Ca, rescaled to $^{40}$Ar by following the same procedure discussed and justified in Ref.~\cite{Martini:2022ebk}, to which we refer the reader for details.
Figure~\ref{fig:microboone_nu} corresponds to Fig.~11 of Ref.~\cite{MicroBooNE:2023cmw}, which includes comparisons with other model predictions and the corresponding $\chi^2$. In this case, the $\chi^2$ obtained with the Martini~\textit{et al.} model is $\chi^2 = 6.4$ for 18 degrees of freedom, which corresponds to a \textit{p} value of 0.994. Among the models considered, the Martini~\textit{et al.} model provides the lowest $\chi^2$ value for this particular measurement. This is likely due to the fact that the 2p2h cross section predicted by our implementation is significantly higher than that predicted by all of the other models considered in Fig.~11 of Ref.~\cite{MicroBooNE:2023cmw}, which are all based on the previous version of the Nieves \textit{et al.} model but GiBUU.\footnote{We notice that in Fig.~11 of Ref.~\cite{MicroBooNE:2023cmw} GiBUU provides the lowest $\chi^2$ ($\chi^2 = 10.2/18$) with respect to the other models considered in that figure. This reinforces our hypothesis that the reason is related to the 2p2h contribution, the GiBUU one being larger than the one of the other Monte Carlos, all based on the previous version of the Nieves \textit{et al.} model, and slightly lower than the predictions of Martini \textit{et al.} for argon, as one can appreciate by comparing Fig.~8 of Ref.~\cite{Mosel:2023zek} with Fig.~5 of Ref.~\cite{Martini:2022ebk}.} Other studies (e.g., Ref.~\cite{Filali:2024vpy}) have indeed shown that more exclusive MicroBooNE measurements indicate a preference for an enhanced 2p2h contribution on argon.

%% file: 5_conclusions.tex
The Martini-Ericson-Chanfray-Marteau RPA-based (anti)neutrino cross-section model has been implemented in the GENIE neutrino event generator. 
Even if this model allows a unified description of several channels (quasielastic, multinucleon emission and the coherent and incoherent single pion production) for both charged- and neutral-current scattering, the present work concerns only the quasielastic and the multinucleon emission in the case of charged-current (anti)neutrino scattering. The implementation of single pion production as well as of electromagnetic and neutral-current scattering is left for future works. 

Basic validations have been performed and presented. All validation steps successfully passed. 

Model predictions in GENIE for $^{12}$C, $^{16}$O, and $^{40}$Ar have been compared with some available experimental measurements from T2K and MicroBooNE. Overall, a reasonable agreement has been observed.

This implementation of the Martini-Ericson-Chanfray-Marteau model in the widely used GENIE event generator is an important step toward eventual reduction of systematic uncertainties due to neutrino-nucleus cross section in future combined analyses of data from the ongoing and next generation precision neutrino experiments.

%% file: 6_acknoledgements.tex
M. M. thanks Magda Ericson, Guy Chanfray, Jacques Marteau and Jean Delorme for all the work that led to the development of the model implemented in the present manuscript. Magda Ericson and Guy Chanfray are also acknowledged by M. M. for the long and continuous collaboration. The authors would also like to acknowledge support and feedback on this work from the T2K and GENIE Collaborations.

%% file: 7_appendix.tex
\appendix
\section{Model predictions for $^{12}$C, $^{16}$O and $^{40}$Ca}
\label{app}

\begin{figure}[h!]
\centering
\includegraphics[width=\linewidth]{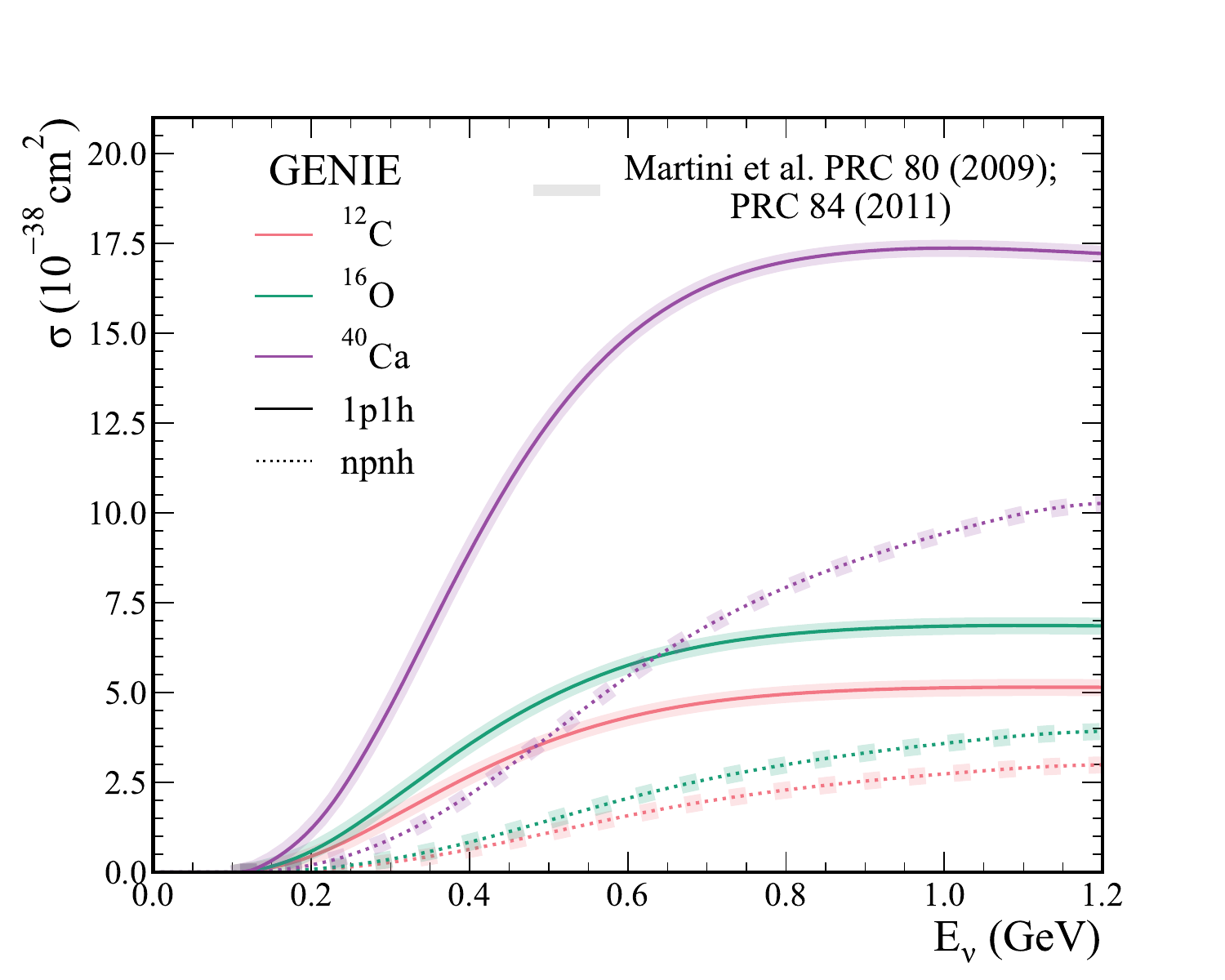}
\caption{
The $\nu_\mu$ charged-current cross section on $^{12}$C (red), $^{16}$O (green), and $^{40}$Ca (violet) as a function of neutrino energy predicted by the model of Martini \textit{et al.} implemented in GENIE. For validation purposes, the results of calculations performed by the authors in Refs.~\cite{Martini:2009uj,Martini:2011wp} are also shown (shaded band).
Different contributions are separately plotted: 1p1h (solid) and npnh (dotted). }
\label{fig:total_nu_alltargets}
 \centering
\includegraphics[width=\linewidth]{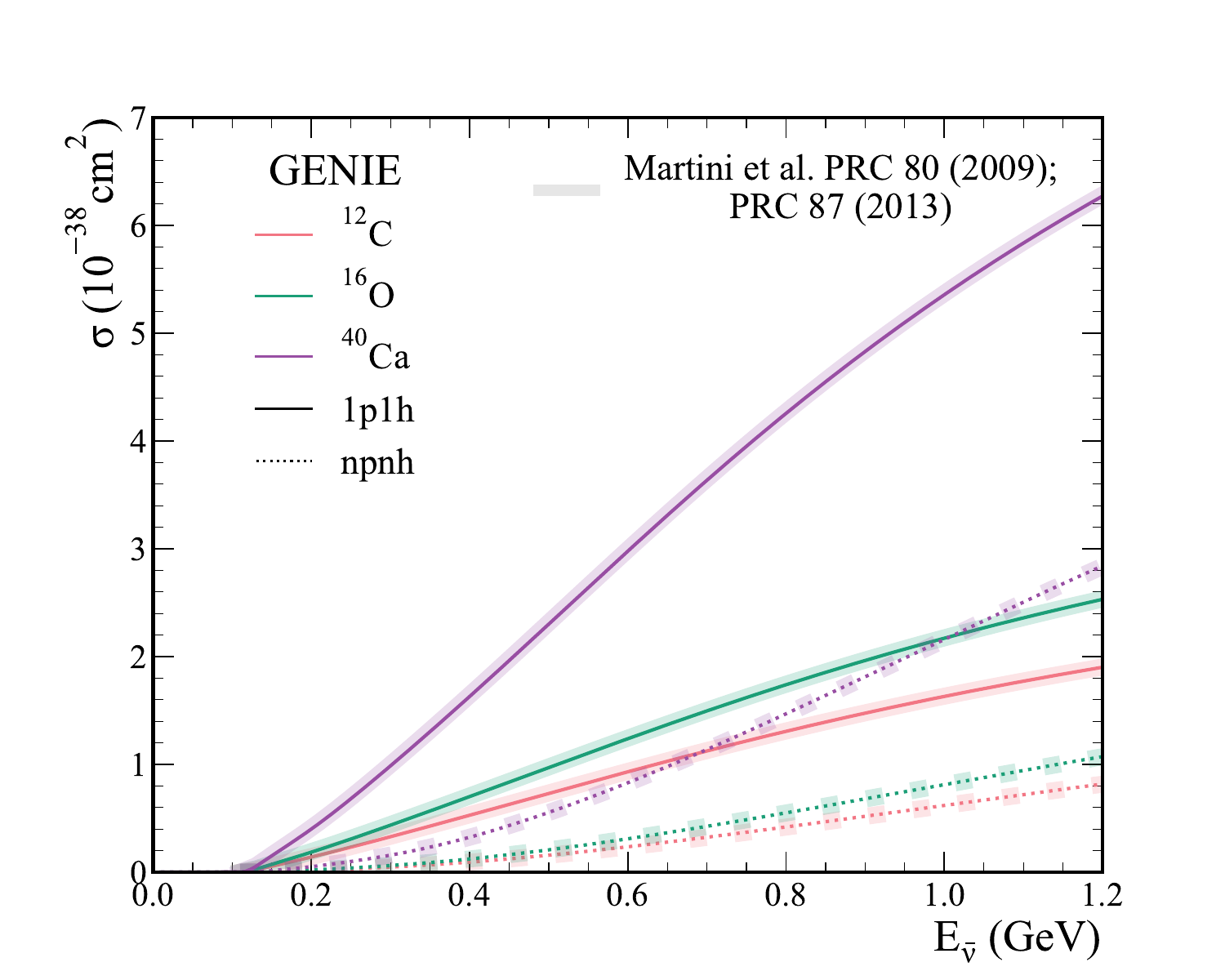}
\caption{
The same as Fig.~\ref{fig:total_nu_alltargets}, but for antineutrinos. In this case, the calculations performed by the authors are the ones of Refs.~\cite{Martini:2009uj,Martini:2013sha}.}
\label{fig:total_anu_alltargets}
\end{figure}

In this appendix we show the model predictions for the $\nu_\mu$ and $\bar \nu_\mu$ charged-current total cross section as a function of the neutrino energy for three different nuclei that are of interest for the ongoing and future experiments: $^{12}$C (already presented and discussed in the main text), $^{16}$O, and $^{40}$Ca (as a proxy for $^{40}$Ar). 
The corresponding comparisons are presented in Figs.~\ref{fig:total_nu_alltargets} and \ref{fig:total_anu_alltargets} for $\nu_\mu$ and $\bar \nu_\mu$, respectively.
A perfect agreement between the GENIE results and the original calculations is observed.





\section{A brief summary of the npnh treatment in the Martini-Ericson-Chanfray-Marteau model}
\label{appendix_npnh}

In this appendix we describe how npnh excitations are treated in the Martini \textit{et al.} model by following an historical perspective which summarizes Sec.~IV C of Ref.~\cite{Martini:2009uj} and Sec.~II of Ref.~\cite{Martini:2010ex}. 

\begin{figure}[th]
\begin{center}
\includegraphics
[width=0.99\columnwidth,clip]{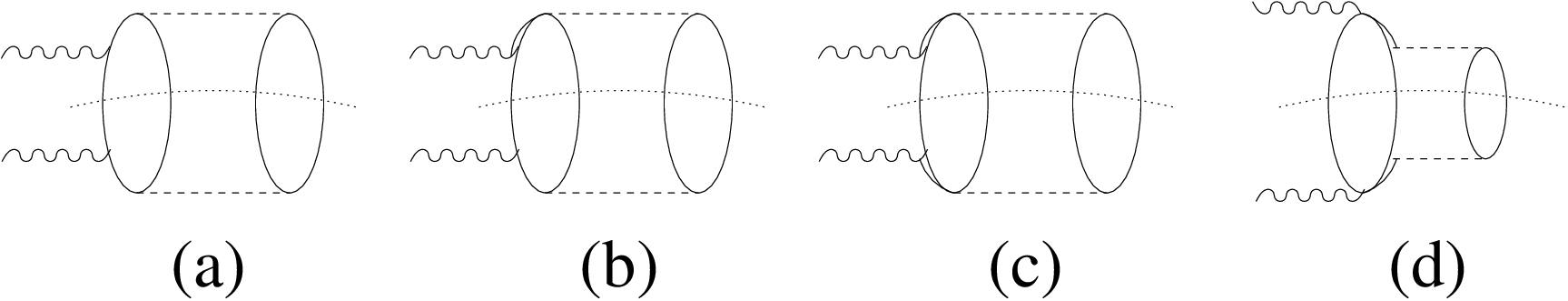}
\end{center}
\caption{\label{fig:bare_2p2h} Examples of Feynman graphs corresponding to 2p2h polarization 
propagators: (a) $NN$ correlations, (b) $N\Delta$ interference, (c), (d) $\Delta\Delta$ MEC. Diagram (d) represents an example of modification of 
the $\Delta$ width in the medium. The wiggled lines represent the external probe, the full lines correspond to the propagation of a nucleon (or a hole), the double lines to the propagation of a $\Delta$ 
and the dashed lines to an effective interaction between nucleons and/or $\Delta$s. The dotted lines show which particles are placed on shell.}
 \end{figure}

There are several sources of npnh excitations. 
One of these sources arises from the modification of the $\Delta$ width in the nuclear medium. This effect was introduced and parametrized by Oset and Salcedo~\cite{Oset:1987re}, which considered the two-body (2p2h) [Fig.~\ref{fig:bare_2p2h}(d)] and the three-body (3p3h) channels in the case of real pion or photon absorption. 
Martini \textit{et al.} have used their parametrization  of the modified $\Delta$ width, although the kinematics of neutrino interaction 
is different since we are in the spacelike region, which could be a source of uncertainty. 
For the other terms not reducible to a modification of a $\Delta$ width, of which some examples are shown in 
Fig.~\ref{fig:bare_2p2h}, they have first used, as in the papers of Marteau \textit{et al.}~\cite{Marteau:1999kt, Marteau:1999jp}, a parametrization of Delorme and Guichon~\cite{DELGUICH} (called ``old'' in Sec.~IV C of Ref.~\cite{Martini:2009uj}). They exploited a calculation by Shimizu and Faessler~\cite{Shimizu:1980kb} 
of the absorptive part of the \textit{p}-wave pion-nucleus optical potential at threshold. It is known that the absorption mechanism of pions is a two-nucleon one, which means 
that in the final state two nucleons are ejected. This is a 2p2h excitation. 
In Ref.~\cite{Shimizu:1980kb}, the absorption is described by three types of terms (see Fig.~\ref{fig:bare_2p2h}).
 The first one [Fig.~\ref{fig:bare_2p2h}(a) with pion lines replacing the weak current ones as in the next diagrams] arises from the $NN$ correlations, essentially from the tensor correlations.
Another one involves a $\Delta$ excitation [Fig.~\ref{fig:bare_2p2h}(c)]. 
The third one is an interference between the nucleon correlation and $\Delta$ terms ($N\Delta$ interference) as in Fig.~\ref{fig:bare_2p2h}(b). The pion absorption calculation of Shimizu and Faessler~\cite{Shimizu:1980kb} is performed  for threshold pion, i.e., for a vanishing three-momentum and an energy $\omega=m_{\pi}$, which does not correspond to the neutrino situation.
Delorme and Guichon~\cite{DELGUICH} have then introduced in each absorption graph the corresponding energy dependence to obtain the  bare 2p2h responses to be inserted in the RPA chain needed for neutrino interaction. The explicit expressions of the corresponding propagators are given in Appendix B~1 of Ref.~\cite{Martini:2009uj}. The shortcoming of these formulas is that the momentum dependence is completely ignored. For this reason in Ref.~\cite{Martini:2009uj}  Martini \textit{et al.} introduced also a different approach that exploits the microscopic calculation of Alberico \textit{et al.}~\cite{Alberico:1983zg} 
specifically aimed at the evaluation of the 2p2h contribution to the isospin spin-transverse response, measured in inclusive $(e,e')$ scattering. Their basic graphs are similar to those of Shimizu and Faessler~\cite{Shimizu:1980kb}. In principle, this way of evaluation is definitely more satisfactory since the kinematical variables are correctly incorporated,
but the results of Alberico \textit{et al.}~\cite{Alberico:1983zg} are available only for a limited set of energy and momenta. 
Martini \textit{et al.} have extended this range to cover the neutrino one through an approximate extrapolation. 
We refer to Ref.~\cite{Martini:2009uj} for the details.
Martini \textit{et al.} hence obtained, 
starting from the calculations of Alberico \textit{et al.}, ``new'' $R^{NN}_{2p-2h}(\omega,q)$ and $R^{N\Delta}_{2p-2h}(\omega,q)$ responses entering in the new $W^{\mu\nu}_{2p2h-NN}$ and $W^{\mu\nu}_{2p2h-N\Delta}$ components of the hadron tensor. For the $\Delta \Delta$ part, which is not well covered in Ref.~\cite{Alberico:1983zg}, Martini \textit{et al.} have kept the previous parametrization, which already presents a proper $q$ dependence owing to the contribution of the in-medium $\Delta$ width~\cite{Oset:1987re}. 
To summarize, in the 2p2h sector, Martini \textit{et al.} include NN correlations, MEC contributions, and NN correlations-MEC interference. Concerning the MEC, they decided to consider only the $\Delta$-MEC and to discard the other contributions from the explicit calculation because these other contributions are peculiar to the external probe and they want a ``universal'' spin-isospin 2p2h response, to use in (and deduced from) different processes, like electron scattering and pion absorption. 
The $\Delta$-MEC contribution is the dominant one in MEC, as shown, for example, in Fig.~9 of~\cite{DePace:2003xu}. 